# Self-Heating Hotspots in Superconducting Nanowires Cooled by Phonon Black-Body Radiation

Andrew Dane[1], Jason Allmaras[2,3], Di Zhu[1], Murat Onen[1], Marco Colangelo[1], Reza Bahgdadi[1], Jean-Luc Tambasco[1], Yukimi Morimoto[1], Ignacio Estay Forno[1], Ilya Charaev[1], Qingyuan Zhao[1], Mikhail Skvortsov[4], Alexander Kozorezov[5], Karl Berggren[1].

[1]Department of Electrical Engineering and Computer Science, Massachusetts Institute of Technology, Cambridge, Massachusetts 02139, USA
[2]Department of Applied Physics and Materials Science, California Institute of Technology, Pasadena, California 91125, USA
[3]Jet Propulsion Laboratory, California Institute of Technology, Pasadena 91109, California, USA
[4]Skolkovo Institute of Science and Technology, Moscow 143026, Russia
[5]Department of Physics, Lancaster University, Lancaster, UK

Controlling thermal transport is important for a range of devices and technologies, from phase change memories to next-generation electronics. This is especially true in nano-scale devices where thermal transport is altered by the influence of surfaces and changes in dimensionality. In superconducting nanowire single-photon detectors, the thermal boundary conductance (TBC) between the nanowire and the substrate it is fabricated on influences most of the performance metrics that make these detectors attractive for applications. This includes the maximum count rate, latency, jitter, and quantum efficiency. Despite its importance, the study of TBC in superconducting nanowire devices has not been done systematically, primarily due to the lack of a straightforward characterization method. Here, we show that simple electrical measurements can be used to estimate the TBC between nanowires and substrates and that these measurements match acoustic mismatch theory across a variety of substrates. Numerical simulations allow us to refine our understanding, however, open questions remain. This work should enable thermal engineering in superconducting nanowire electronics and cryogenic detectors for improved device performance.

Superconducting nanowires are the basis for a number of quantum technologies including a variety of single-photon detecting devices [1] and quantum phase-slip junctions [2], however the study of their thermal properties is often incidental. In phase slip junctions, heating is detrimental to the coherent tunneling of phase slips, though it may enable the observation of single macroscopic quantum tunneling events [3]. Superconducting nanowire single photon detectors (SNSPDs) [1], on the other hand rely on the creation of localized, photo-induced normal regions, or hotspots, in order to detect infrared photons [4][5][6]. Energy deposited into the SNSPD is eventually released into the substrate in the form of phonons. Macroscopically, the net rate of phonon emission is quantified by the thermal boundary conductance (TBC) between the wire and substrate.

The TBC between SNSPDs and dielectric substrates is one of the main determinants of maximum (non-latching) device speed [5], it is thought to affect quantum efficiency [7], jitter and latency [8], and it may affect observed switching currents, and dark count rates. At the early stages of photo-detection, the energy from a photon absorbed in an SNSPD is divided among a small number of quasiparticle and phonon excitations [9]. Pair-breaking phonons that escape into the substrate reduce the energy available to disrupt the superconducting state. Thus, a lower TBC may increase device detection efficiency [7]. On the other hand, pair breaking phonons reflected at the substrate interface could lead to switching of the device at a later time, increasing the latency and jitter of the device [8]. After hotspot formation, electro-thermal feedback on its growth determines the conditions under which a device can operate in a free running mode, or whether it will latch into a resistive state; faster detectors require increased TBC [5]. Finally, as device DC bias currents are ramped up to increase detection efficiency and lower jitter, eventually vortices will be drawn across the wire [10]. Vortex crossings release energy which can lead to localized heating and increased vortex flow, and at sufficiently high currents, thermal runaway and hotspot formation [3]. Increased cooling could help reduce dark counts at a given DC bias. At sufficiently high dark count rates, the wire switches to a stable resistive state, well before the theoretical depairing current [11]. Seemingly all SNSPD performance metrics relate to how quickly heat is removed from the nanowire. Despite this, the TBC between SNSPDs and substrates has only been studied in a handful of cases [12][13].

Here, we attempt to quantify the TBC between superconducting nanowires and substrates using measurements of the current needed to sustain a hotspot inside the nanowire, known as the self-heating hotspot current ($I_{hs}$) [14]. Measurements of this type have been attempted previously [4], though usually with micrometer scale devices, one substrate type, and without clear comparisons to theoretical expectations [15][16]. In order to measure the TBC between nanowires and substrates, and clarify the description of hotspot in nanowires, we compare measurements of $I_{hs}(T_b)$ for 17 NbN nanowires across 6 different substrate materials, with analytical calculations and finite element electro-thermal simulations.

In order to make sense of our experimental results, we first review a theoretical model used extensively to interpret similar measurements. The Skopkol Beasley Tinkham (SBT) model [17], including later generalizations [18][19][20], forms the starting point for understanding hotspots in superconducting nanowires. The model considers a one-dimensional normal domain inside of an otherwise superconducting wire, centered at $x = 0$ and extending to $\pm x_N$. The heat equation in the normal and superconducting regions is given by:

$$-(K_N \cdot w^2 d)\frac{\partial^2 T}{\partial x^2} + \beta w^2 (T^n - T_b^n) = I^2 R_n \qquad (|x| < x_N) \qquad (1a)$$

$$-(K_S \cdot w^2 d)\frac{\partial^2 T}{\partial x^2} + \beta w^2 (T^n - T_b^n) = 0 \qquad (|x| > x_N) \qquad (1b)$$

where $K_N$ and $K_S$ are the thermal conductivities of the wire in the normal and superconducting regions. $w$ and $d$ are the width and thickness of the wire, $I$ is the current, $R_n$ is the sheet resistance in the normal state, $\beta$ is a generic thermal boundary conductance with units of W/m²K$^n$, and $T_b$ is the bath/substrate temperature. The temperature at $\pm x_N$ is assumed to be $T_c$. The temperature profile along the wire, $T(x)$, is solved for such that the heat flow at $\pm x_N$ is continuous, and $T \to T_b$ as $x \to \pm\infty$. For a given $x_N$, the solution to Eq. (1) determines a current-voltage pair. By solving for a range of $x_N$, we can trace out a current-voltage (IV) curve that contains a distinct region of near-constant current for a range of voltages. This current is the hotspot current, $I_{hs}$.

When the hotspot is sufficiently long, we expect that the temperature around $x = 0$ will be nearly constant, such that $\frac{\partial^2 T}{\partial x^2} \approx \frac{\partial T}{\partial x} \approx 0$ [19], allowing us to drop the first term on the left hand side of (1a):

$$\beta w^2 (T_{hs}^n - T_b^n) = I_{hs}^2 R_n \qquad (2)$$

Here, $T_{hs}$ is the hotspot temperature, the nearly constant temperature in the normal domain near $x = 0$. $I_{hs}$ is the current required to maintain the hotspot via Joule heating. This relation was found by Dharmadurai and Satya Murthy (DSM) [19], to outperform their own more sophisticated attempts to model $I_{hs}$ data for long superconducting wires with ~mm widths and 10-20 nm thicknesses, originally measured in ref [18]. Interestingly, these were Al samples evaporated in an oxygen atmosphere, with resistivities in the range of recently reported granular aluminum [21]. DSM assumed that $T_{hs} = T_c$ for all $T_b$. This assumption is not compatible with (1), which predicts that $T_{hs}$ is a function of $T_b$, with $T_{hs}(T_b = 0) = T_c(1 + 1/n)$ [17][18][19].

The value of the exponent $n$ in the second term on the left-hand side of (1) is an important parameter that captures the relevant physics and dimensionality of the excitations which cool the hotspot. For instance, for clean bulk metals, $n = 5$ would be appropriate if the bottleneck to heat flow was between electrons and phonons within the wire [22][23] while $n = 4$ could describe the same for a clean metal membrane [24]. For highly disordered bulk metals, $n = 6$ for electron-phonon coupling has been derived [25]. Non-integer values of $n$ may occur due to the roughness or structure of the interface that makes transmission wavelength-dependent [22][26]. Importantly, when the difference between the hotspot temperature and the bath is small, this term can be linearized to a term of the form $n\beta T_b^{n-1} w^2 (T - T_b)$, as is often done [27]. The additional terms in the prefactor are combined into a linear heat transfer coefficient, $\alpha = n\beta T_b^{n-1}$.

The original SBT paper [17] used a linearized TBC which was quickly shown to be invalid at low $T_b$ [18]. The SBT model with linear TBC predicts that $I_{hs}(T_b) \propto \sqrt{T_c - T_b}$. This prediction has been used in the past to fit $I_{hs}(T_b)$ data to estimate $\alpha$ [16]. When we square the hotspot current predicted by the original SBT, we find it is linearly related to the bath temperature, $I_{hs}^2(T_b) \propto (T_c - T_b)$. Written in this way, the deficiency in the original SBT model is made clear, when compared with experimental results [18][19]. However, this linearization was incorporated into initial descriptions of SNSPDs [1], without being addressed as such.

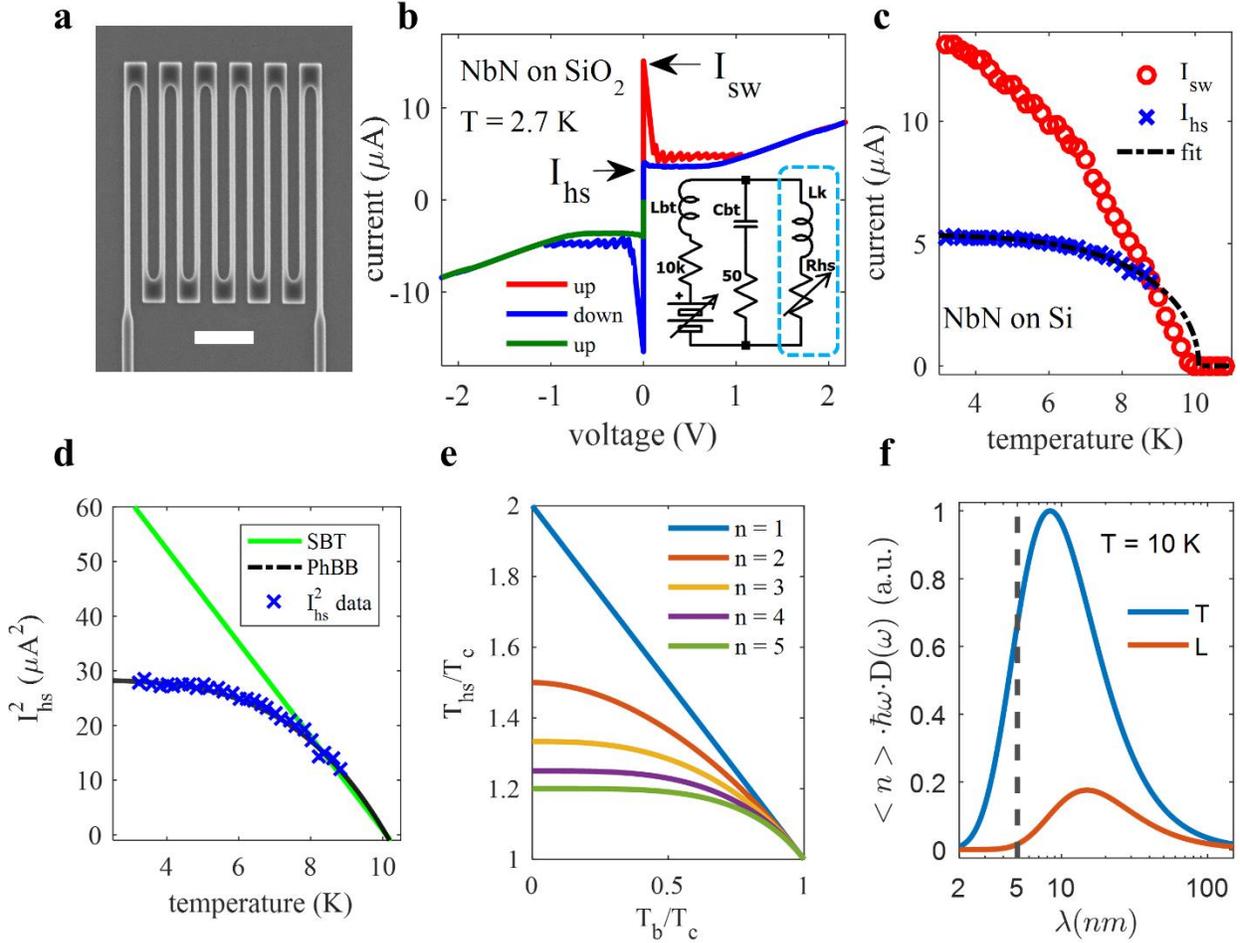

**Figure 1. Measurements and modelling of $I_{hs}(T_b)$.** (a) Scanning electron micrograph of a device measured for this work. Scale bar is one µm. (b) Typical hysteretic IV curve of a superconducting NbN nanowire with the switching ($I_{sw}$) and hotspot ($I_{hs}$) currents labelled. The measurement circuit is inset. Only the nanowire was cooled (blue dashed box). (c) $I_{sw}(T_b)$ and $I_{hs}(T_b)$ for NbN on Si device. Fitting of $I_{hs}$ is shown and discussed below. (d) The $I_{hs}(T_b)$ data and fit from (c) was squared and replotted. Displayed this way, it is easy to see the failure of the linearized SBT model to capture the shape of the data. (d) Calculated hotspot temperature ($T_{hs}$) as a function of $T_b$ and the exponent $n$ which describes the power law cooling to the substrate. (e) Normalized energy density carried by three-dimensional longitudinal (orange) and transverse (blue) phonons in NbN as a function of wavelength at 10 K, using the Debye model for the phonon density of states. Both curves have been normalized by the maximum of the transverse curve. Typical film thickness (5 nm) indicated by dashed line.

Figure 1 shows an example of our DC electrical measurements of $I_{hs}$ at various $T_b$, which we use to illustrate that the linearized SBT model is not consistent with experiment, and points out open questions related to properly explaining $I_{hs}$. IV curves like the one shown in Fig. 1(a) were measured vs bath temperature ($T_b$) for 17 NbN nanowires across 6 different substrate materials. $I_{hs}$ is identified as the constant current 'plateau' [1] at non-zero voltages, observed when ramping the bias down after switching. Unlike the switching current ($I_{sw}$), $I_{hs}$ is deterministic and immune to noise in this configuration [28]. The measurement circuit is shown inset. IV curves were measured at 0.2 K increments of $T_b$, up to 12 K, and the resulting $I_{hs}(T_b)$ and $I_{sw}(T_b)$ for one device on Si is plotted in 1(b). When $I_{sw}(T_b) > I_{hs}(T_b)$, the IV curve is hysteretic [29], and while non-hysteretic IV curves may contain evidence of hotspot formation [30], we limit our identification of $I_{hs}$ to temperatures where the IV curve is hysteretic and the presence of the hotspot unambiguous. If we square the $I_{hs}(T_b)$ data from 1(b), we can immediately see that it is inconsistent with the predictions of the linearized SBT model, as shown in Fig. 1(c). In Fig. 1(d) we plot $T_{hs}(T_b)$ at the center of a long nanowire numerically calculated using Eq. (1) with $K_N = K_S$ for $n = 1 - 5$, to illustrate a potential issue with assuming that $T_{hs} = T_c$ for all $T_b$, as done by DSM. Lastly, in Fig. 1(e) we plot the expected energy content for phonon modes in NbN at 10 K vs the wavelength of that mode, assuming that the NbN is three-dimensional. The vast majority of the energy in such a system is carried by phonons with wavelengths exceeding the 5 nm film thickness typical for SNSPDs, implying that the correct description of the phonon system is not the simple three-dimensional picture. Thus, we might expect to need to modify both the description of the electron-phonon coupling inside of the wire [31], as well as the phonon-phonon coupling across the material boundaries [32].

All of our hotspot current data can be explained by a thermal boundary conductance that is determined by phonon black body radiation into the substrate, and as a result, the phonon-emissivity between the nanowire and substrate is the primary factor in determining $I_{hs}(T_b)$. Equation (2) with $n = 4$, $I_{hs}^2 R_n = \beta w^2(T_{hs}^4 - T_b^4)$, equates the Joule heating in the hotspot with the net heat flow out of the wire due to a detailed balance of phonon flux at the interface. This value of $n$ is appropriate for three-dimensional phonons, and despite our suggestion that the thickness of the wire should alter the dimensionality, we find that we can fit all of our data using Eq. (2) with $n = 4$, and that the fitting parameters match calculated values. The right-hand side has the same form as the net radiative heat transfer from a black body at temperature $T_1$ to another at $T_2$ due to the Stephan-Boltzman law for blackbody radiation of photons, $P_{1 \to 2} = \sigma \epsilon A(T_1^4 - T_2^4)$, where A is an effective surface area, $\epsilon \leq 1$ is an effective emissivity, and $\sigma$ is the usual Stephan-Boltzmann constant. In the phonon case, $\sigma$ becomes $\sigma_{ph} = \frac{\pi^2 k_B^4}{120 \hbar^3}\left(\frac{1}{c_{L1}^2} + \frac{2}{c_{T1}^2}\right)$ [27], where $c_{L1}$ and $c_{T1}$ are the longitudinal and transverse speeds of sound in one of the materials. For an average speed of sound, such that $\left(\frac{1}{c_{L1}^2} + \frac{2}{c_{T1}^2}\right) = \frac{3}{c_{avg}^2}$, we can write $\sigma_{ph} = \frac{\pi^2 k_B^4}{40 \hbar^3 c_{avg}^2}$ as done in [33]. Note that this description is the same as the well-known Kapitza resistance. However, 'Kaptiza resistance' is often used ambiguously, or in situations where small temperature differences invite linearization. Here, the linearization does not work, as shown in Fig. 1(c). We therefore avoid 'Kaptiza resistance' in favor of 'phonon black body radiation' similar to Ref. [33].

In order to compare experimental data with theoretical expectations, we calculated the phonon-emissivity between NbN and various substrate materials, using the acoustic and diffuse mismatch models. In the acoustic mismatch model (AMM) the interface between materials is assumed to be ideal, and incident phonons are transmitted and reflected specularly, in proportions that satisfy continuum mechanics at the interface [34]. In the opposite limit, the diffuse mismatch model (DMM) assumes that the interface is completely diffusive; incident phonons lose all memory of the direction of propagation before impinging on the interface. Phonons that arrive at the interface are scattered to one or the other side in proportion to the phonon density of states on either side of the interface. Mathematically, the emissivity values calculated using these two models are:

$$\epsilon_{AMM} = \left(\frac{\eta_L}{c_{L1}^2} + \frac{2\eta_T}{c_{T1}^2}\right)\left(\frac{1}{c_{L1}^2} + \frac{2}{c_{T1}^2}\right)^{-1}$$

$$\epsilon_{DMM} = \left(\frac{1}{c_{L2}^2} + \frac{2}{c_{T2}^2}\right)\left(\frac{1}{c_{L1}^2} + \frac{2}{c_{T1}^2} + \frac{1}{c_{L2}^2} + \frac{2}{c_{T2}^2}\right)^{-1}$$

$0 \leq \eta_L \leq 1$ and $0 \leq \eta_T \leq 1$ are the angle-averaged transmission factors for phonons originating in material 1, incident on the interface with material 2, for longitudinal and transverse phonons. We calculated $\eta_i$ using literature values for material properties (see Table S4) while accounting for possible total internal reflection, as described by Kaplan [35]. These calculations are discussed in detail in appendix A of [36]. We verified our AMM and DMM calculations by recreating table II from [27]. In addition, we tabulated values of $\sigma_{ph}\epsilon_{AMM}$ for a variety of superconductor-substrate pairs that may be useful for SNSPD applications. The results are given in Table S2, and S5 respectively.

In Fig. 2, we summarize our fitting of $I_{hs}(T_b)$ data using Eq. (2), and compare the extracted phonon emissivity to the predictions of AMM and DMM. We focus primarily on the case of $n = 4$, using $\beta$ and $T_{hs}$ as fitting parameters. In Fig. 2(a), a representative subset of the $I_{hs}(T_b)$ data and fit lines is shown. Measured values of $R_n$ at 12 K, and measured or design values of $w$ were input into the fitting expression. All but one fit had $R^2 > 98\%$. The $R^2$ for NbN on polyethylene terephthalate was 94%. Data and fit lines from Fig. 2(a) are re-plotted in 2(b) in order to more easily compare wires of different width and $T_c$ value and more clearly illustrate the observed range of $\beta_{\text{fit}}$ values. Figure 2(c) and 2(d) are histograms of the fitting parameter $\beta_{\text{fit}}$ divided by calculated values of $\sigma_{\text{ph}}\epsilon_{\text{XMM}}$. Entries into the histogram are colored according to the substrate material. In these histograms, we have excluded the two NbN-Ti bilayer on SiO$_2$ devices, which appear to have partially alloyed during the fabrication process, making it difficult to describe them acoustically (see Fig. S3). While the values of $\beta_{\text{fit}}$ appear systematically lower than $\sigma_{\text{ph}}\epsilon_{\text{AMM}}$, our data suggests that AMM captures the essential physics better than DMM. This is consistent with recent work where the value of the thermal boundary conductance consistent with measurements was about 20% of the value calculated using DMM [37]. This analysis also worked for data extracted from [16], which erroneously followed the linearized SBT model, yielding $\beta_{\text{fit}}/\sigma_{\text{ph}}\epsilon_{\text{AMM}} = 0.72$ (see Fig. S2).

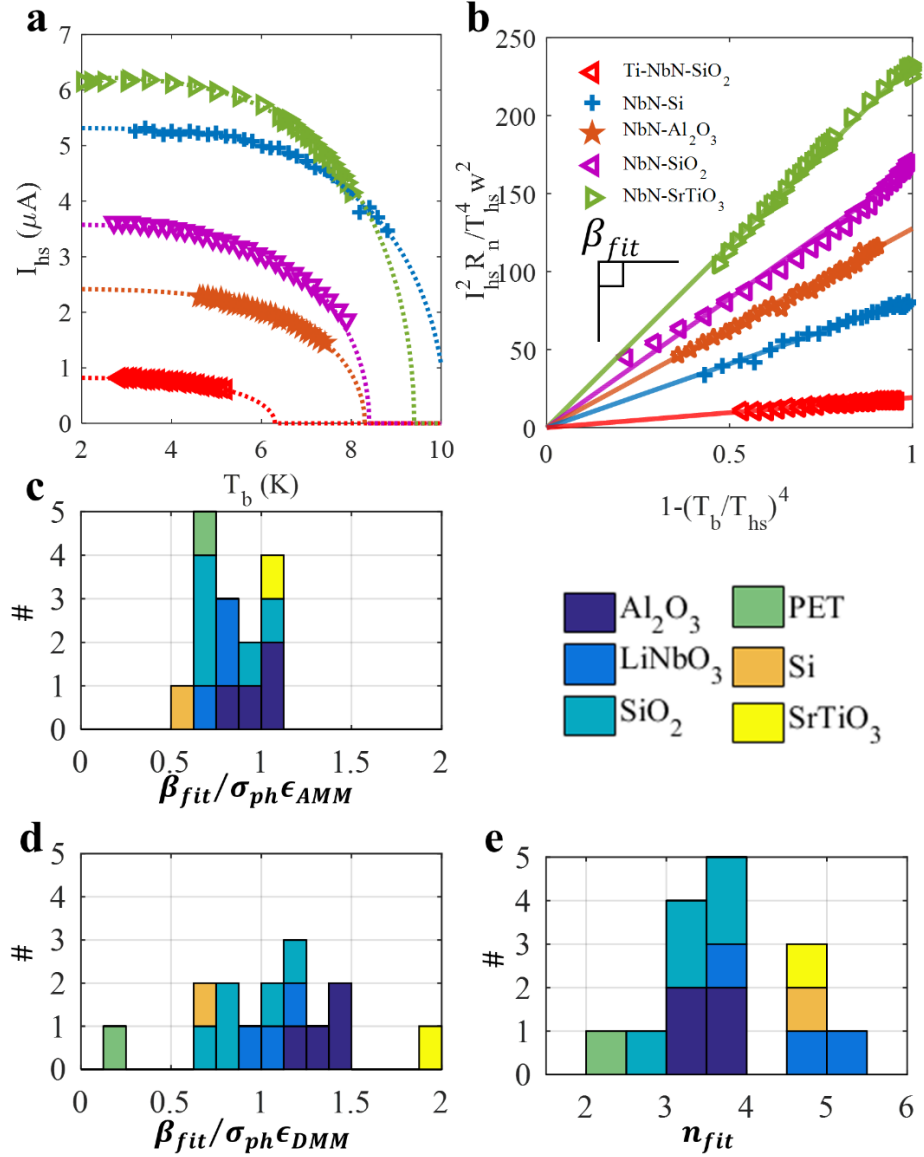

**Figure 2. Comparing the extracted phonon emissivity to AMM and DMM.** (a) $I_{hs}(T_b)$ for representative nanowires. Equation (2) is fit to the data using $\beta$ and $T_{hs}$ as fitting parameters with $n = 4$. Legend is shared with (b). (b) Transformation of data in (a) allowing us to compare wires with different widths and fitted $T_{hs}$. Plotted this way, the fit curves become lines with slope $\beta_{fit}$. (c-d) each histogram entry is colored to indicate substrate. (c) Histogram of $\beta_{fit}/(\sigma_{ph}\epsilon_{AMM})$. (d) Histogram of $\beta_{fit}/(\sigma_{ph}\epsilon_{DMM})$. (e) Histogram of $n_{fit}$ that results from refitting all $I_{hs}(T_b)$ while allowing $\beta$, $T_{hs}$, and $n$ to vary.

Deviations from the fit shape were noticeable in some data sets. In Fig. 2(a)(b), for instance, in our NbN on SiO$_2$ data (purple triangles) and in our NbN on polyethylene terephthalate (not shown; to be discussed at length elsewhere). The substrate we call SiO$_2$ was a 300 nm thick silicon thermal oxide layer on top of Si. The additional phonon scattering from the SiO$_2$-Si boundary, as well as potential anomalously high intrinsic scattering in the SiO$_2$ [13], may push these samples away from the phonon radiation regime and cause the shape deviation observed. This may also explain why we predict (measure) a higher $\sigma_{\text{ph}}\epsilon_{\text{AMM}}(\beta)$ for NbN on SiO$_2$ than Al$_2$O$_3$, and yet NbN SNSPD count rate measurements suggest higher TBC on Al$_2$O$_3$ [38]. Prompted by these deviations in shape, we repeated all fits, allowing $n$ to vary in addition to $\beta$ and $T_{\text{hs}}$. The resulting values of $n$ are plotted as a histogram in 2(e). While this histogram gives the impression that a number of our samples may be described by a cooling mechanism with $n = 5$, we find from our simulation work that the process of extracting $n$ from $I_{\text{hs}}(T_{\text{b}})$ data may be unreliable.

To support and extend our analysis, we used a one-dimensional finite element simulation to repeat the present experiments in-silico. This simulation primarily follows [6] and consists of a set of coupled heat balance equations that determine the electron ($T_{\text{e}}(x)$) and phonon ($T_{\text{ph}}(x)$) temperatures inside the hotspot, incorporating effects of the local superconducting gap $\Delta(T_{\text{e}}(x))$. Allowing for two temperatures inside of the hotspot is in contrast to previous analytical models mentioned so far and our foregoing discussion, which tacitly assumed $T_{\text{e}} = T_{\text{ph}} = T_{\text{hs}}$. The temperature of the electron subsystem at every point in the simulation is governed by heat diffusion among electrons, Joule heating due to the hotspot current, and coupling between electrons and phonons proportional to $T_{\text{e}}^5 - T_{\text{ph}}^5$. Similarly, the local temperature of the phonon system is a balance between phononic heat conduction along the wire, heat received from the electron subsystem, and heat that escapes to the substrate described by a term proportional to $T_{\text{ph}}^4 - T_{\text{b}}^4$.

The electron-phonon coupling strength can be quantified by a characteristic timescale for energy exchange, given by $\tau_{ep}$, and is determined by material properties and geometry. Similarly, the coupling time for phonons in the wire and phonons in the substrate is given by $\tau_{pp}$ and is related to the phonon emissivity and relative heat capacities of the two subsystems. Lastly, the boundaries of the hotspot separate the superconducting and normal phases, and this boundary is dependent on both temperature and current [39]. To account for this in our model, the local critical temperature is reduced by the presence of current; we invert the Bardeen equation for the depairing current as a function of temperature, to give us an equation for the $T_c$ as a function of current. Further details of the model are given in the supplementary materials.

Figure 3 summarizes the results of our simulation efforts which included: (1) calculating the temperatures of the electron ($T_e$) and phonon ($T_{ph}$) sub-systems in the center of a stable hotspot, (2) recreating the measurement and fitting procedure with simulated data. Figure 3(a) illustrates the conceptual difference between how our fitting equation treated the hotspot (left), and how the simulation did (right), with red arrows indicating the net power flow. In 3(b) we report the simulated $T_e$, $T_{ph}$, and the effective $T_c$ in the center of a simulated hotspot, as $T_b$ is varied. As $T_b$ is reduced, $T_e$ increases in a manner qualitatively similar to what SBT predicts for $T_{hs}$ (see Fig. 1(d)), with a reduced magnitude. $T_{ph}$ is bounded above by $T_e$ and below by $T_b$, but it's exact value depends on the relative strengths of the electron-phonon coupling in NbN, and the phonon-phonon coupling across the wire-substrate boundary. For NbN, we find that $T_{ph}$ is nearly constant across $T_b$. This helps explain why we can use one temperature to fit our $I_{hs}$ data. In Fig. 3(c) we fit the simulated $I_{hs}(T_b)$ data with the phonon blackbody radiation expression. When we set $n = 4$, and allow $\beta$ and $T_{hs}$ to vary, the best fit value of $\beta$, $\beta_{fit} = 528$ W/m²K⁴, is within 2% of the value entered into the simulation beforehand, $\beta_{sim} = 518$ W/m²K⁴. When we allow $\beta$, $n$ and $T_{hs}$ to vary, the best fit $n$ is close to 3.5. Based on these simulation results, our fitting procedure seems unreliable in determining $n$.

In order to better understand how much the material properties of the film and substrate might influence the accuracy of our method, we repeated our simulation and fitting procedure while varying the strength of the electron-phonon coupling in the wire and the thermal boundary conductance to the substrate. The results are reported in Fig. 3(d). Since our model was originally setup for time-dependent simulations of photo-detection, the natural way for us to vary the coupling strengths was to vary the relevant time scales that enter the simulation. By varying the characteristic time, $\tilde{\tau}_0$ [40], as well as a scaled phonon escape time, $\gamma_{ph}\tilde{\tau}_{esc}$, we could probe different electron-phonon coupling strengths and thermal boundary conductances, respectively. This method follows [6], with further details given in the SI. In Fig. 3(d) we plot a heat map of $\beta_{fit}/\beta_{sim}$ vs $\tilde{\tau}_0$ and $\gamma_{ph}\tilde{\tau}_{esc}$. The red ellipse approximates the region where we would expect to find our NbN, based on published and measured data. In this region $\beta_{fit}/\beta_{sim}$ takes on values that are similar to what we found experimentally (see Fig. 2(c)), roughly from 0.4 to 1.2. We take this as an indication that our simple measurement and fitting procedure does as well as theoretically possible, and that when combined with knowledge of the electron-phonon coupling strength in a given superconducting material, could be used to improve the accuracy of this method.

Despite decades of progress, the acoustic and diffuse mismatch models often fail to accurately predict the observed thermal boundary conductance, which is why we consider our results to be highly encouraging. An agreement of ~20% between measurement and calculation is looked upon favorably [37]. In comparison, the values we extract experimentally are significantly closer to the values expected from AMM, despite using literature values of acoustic properties for all calculations. The reason for this agreement is still unclear. It may be that the stringent requirements put on the material in order to form state of the art SNSPDs also improves their performance in this application. High performance SNSPDs are fabricated from material that is flat, smooth, and without significant variation in the material properties along the wire. Sputtering ensures that the material is well bonded to the substrate [41]. The result may be a device that is ideally suited for nanoscale thermal transport measurements of this kind. Of course, there is still the open question about the degree of phonon localization in the wire, and even more basic questions about the description of electron-phonon coupling in nanoscale dirty metals [42]. Others have reported a discrepancy between the expected phonon dimensionality, and that which best described heat transfer in short and thin superconducting nanowires [39]. However, the good match between our experimental results, analytic calculations and simulated results suggest that the effect is small.

In summary, our method of extracting $\beta$ is simple, the extracted values match those expected by acoustic modelling to a surprising degree, and our electro-thermal simulations give us a better understanding of the circumstances under which better accuracy can be expected. Thus, we believe that superconducting nanowires such as those being created for high efficiency single-photon detection can be a convenient platform for probing nanoscale heat transfer phenomena.

This work should pave a path towards improving superconducting device performance by controlling the ease with which phonons can escape the device after excitation. For NbN SNSPDs, reducing $\beta$ by choice of substrate could lead to improved detection efficiency at longer photon wavelengths [7]. Increasing $\beta$ has already been shown as a promising route for improving the bandwidth of NbN hot electron bolometers [43], and phonon trapping has been shown to improve single-photon energy resolution in kinetic inductance detectors [44]. The same would improve SNSPD count rates [5]. The well-defined temperature in self-heating hotspots formed in superconducting structures should lend itself to the use of phononic structures to increase or decrease transmission at appropriate wavelengths. An in-plane phononic crystal approach has been proposed in order to extend the effective quasiparticle lifetimes in kinetic inductance detectors (KIDs) by patterning holes in a membrane supporting the KID [45]. Lastly, superconducting quantum circuits made from aluminum may benefit from a choice of substrate with better acoustic matching. Aluminum is poorly acoustically matched to commonly used substrates, such as silicon and sapphire [35]. This may increase the effective lifetime of quasiparticles in the device since pair breaking phonons created by recombination have a relatively low chance of escaping to the substrate without being reflected at the interface.

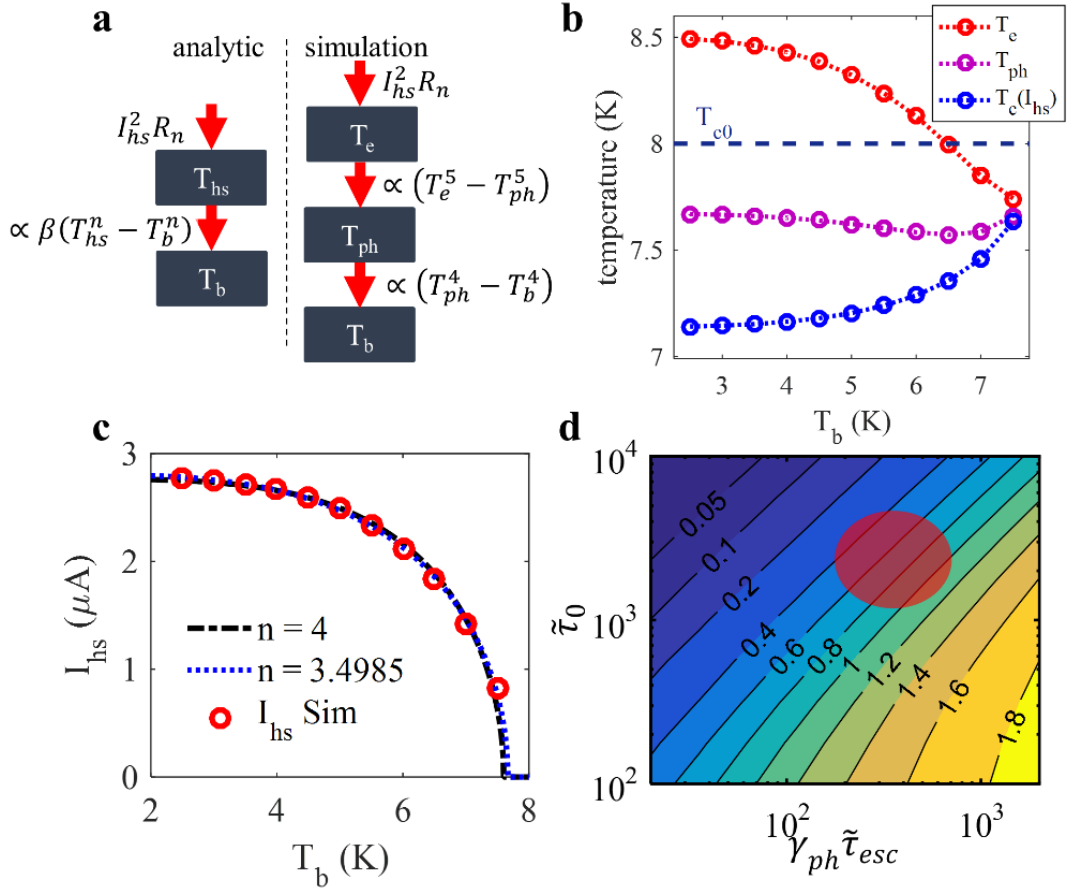

**Figure 3. Electro-thermal simulations of hotspots in superconducting nanowires.** (a) Illustration of the conceptual difference between the analytic model, and numerical simulation. (b) Simulated electron ($T_e$) and phonon ($T_{ph}$) temperatures in the center of a hotspot, as a function of $T_b$, along with $T_c(I_{hs})$. (c) Simulated $I_{hs}(T_b)$ (red circles), along with fits using Eq. (2) with $n = 4$ (black), and $n$ as free parameter (blue). (d) Heat map of $\beta_{fit}/\beta_{sim}$ as a function of a characteristic electron-phonon coupling time ($\tilde{\tau}_0$), and a scaled phonon escape time ($\gamma_{ph}\tilde{\tau}_{esc}$). The red circle indicates where we expect the values for NbN to fall based on literature and our measurements.

Acknowledgements

This work was supported by the Skoltech Next Generation Program grant no. 1921/R, as well as the DARPA Detect Project, grant no. DARPA ARO W911NF1620192. A.D. and J.A. were each supported by a NASA Space technology research fellowship, grant no.s NNX14AL48H and NNX16AM54H respectively. The authors would like to thank Jim Daley and Mark Mondol who enabled our use of the Nanostructures Laboratory at MIT where the majority of the sample fabrication took place. We would also like to thank Donnie Keathley and Navid Abedzadeh for their help reviewing drafts of this manuscript, as well as all other members of the Quantum Nanostructures and Nanofabrication Group at MIT.

Author Contributions

A.D. and K.B conceptualized these experiments. Fabrication of devices was done by A.D, M.O., M.C., R.B., J.T., Y.M. and I.F. Measurements were performed by A.D, D.Z., M.O., M.C., and Q.Z. Analysis and interpretation of the data was done by A.D., J.A, I.C., M.S., A.K., and K.B. Numerical simulations were performed by J.A with input from A.K. K.B. supported and supervised this work. A.D. prepared the manuscript with input from all co-authors.

Competing Interests.

The authors declare no competing interests.

References:

[1]     G. N. Gol'tsman *et al.*, "Picosecond superconducting single-photon optical detector," *Appl. Phys. Lett.*, vol. 79, no. 6, p. 705, 2001.

[2]     J. E. Mooij and Y. V. Nazarov, "Superconducting nanowires as quantum phase-slip junctions," *Nat. Phys.*, vol. 2, no. 3, pp. 169–172, Feb. 2006.

[3]     M. Sahu *et al.*, "Individual topological tunnelling events of a quantum field probed through their macroscopic consequences," *Nat. Phys.*, vol. 5, no. 7, pp. 503–508, May 2009.

[4]     J. K. W. Yang, A. J. Kerman, E. A. Dauler, V. Anant, K. M. Rosfjord, and K. K. Berggren, "Modeling the Electrical and Thermal Response of Superconducting Nanowire Single-Photon Detectors," *IEEE Trans. Appl. Supercond.*, vol. 17, no. 2, pp. 581–585, Jun. 2007.

[5]     A. J. Kerman, J. K. W. Yang, R. J. Molnar, E. A. Dauler, and K. K. Berggren, "Electrothermal feedback in superconducting nanowire single-photon detectors," *Phys. Rev. B*, vol. 79, no. 10, p. 100509, Mar. 2009.

[6]     D. Y. Vodolazov, "Single-Photon Detection by a Dirty Current-Carrying Superconducting Strip Based on the Kinetic-Equation Approach," *Phys. Rev. Appl.*, vol. 7, no. 3, pp. 1–19, 2017.

[7]     Y. Ota, K. Kobayashi, M. Machida, T. Koyama, and F. Nori, "Full Numerical Simulations of Dynamical Response in Superconducting Single-Photon Detectors," *IEEE Trans. Appl. Supercond.*, vol. 23, no. 3, Jun. 2013.

[8]  J. P. Allmaras, A. G. Kozorezov, B. A. Korzh, K. K. Berggren, and M. D. Shaw, "Intrinsic Timing Jitter and Latency in Superconducting Nanowire Single-photon Detectors," *Phys. Rev. Appl.*, vol. 11, no. 3, p. 034062, 2019.

[9]  A. G. Kozorezov *et al.*, "Fano fluctuations in superconducting-nanowire single-photon detectors," *Phys. Rev. B*, vol. 96, no. 054507, pp. 1–13, 2017.

[10] L. N. Bulaevskii, M. J. Graf, C. D. Batista, and V. G. Kogan, "Vortex-induced dissipation in narrow current-biased thin-film superconducting strips," *Phys. Rev. B - Condens. Matter Mater. Phys.*, vol. 83, no. 14, pp. 1–9, 2011.

[11] S. Frasca *et al.*, "Determining the depairing current in superconducting nanowire single-photon detectors," *Phys. Rev. B*, vol. 100, no. 5, pp. 1–7, 2019.

[12] E. M. Baeva *et al.*, "Thermal Properties of NbN Single-Photon Detectors," *Phys. Rev. Appl.*, vol. 10, no. 6, p. 1, 2018.

[13] J. P. Allmaras, A. G. Kozorezov, A. D. Beyer, F. Marsili, R. M. Briggs, and M. D. Shaw, "Thin-Film Thermal Conductivity Measurements Using Superconducting Nanowires," *J. Low Temp. Phys.*, pp. 1–7, 2018.

[14] A. Gurevich and R. G. Mints, "Self-heating in normal metals and superconductors," *Rev. Mod. Phys.*, vol. 59, no. 4, 1987.

[15] A. Stockhausen *et al.*, "Adjustment of self-heating in long superconducting thin film NbN microbridges," *Supercond. Sci. Technol.*, vol. 25, no. 3, 2012.

[16] M. W. Johnson, A. M. Herr, and A. M. Kadin, "Bolometric and nonbolometric infrared photoresponses in ultrathin superconducting NbN films," *J. Appl. Phys.*, vol. 79, no. 9, pp.

7069–7074, 1996.

[17] W. J. Skocpol, M. R. Beasley, and M. Tinkham, "Self-heating hotspots in superconducting thin-film microbridges," *J. Appl. Phys.*, vol. 45, pp. 4054–4066, 1974.

[18] S. Yamasaki and T. Aomine, "Self-heating effects in long superconducting thin films over a wide temperature range," *Jpn. J. Appl. Phys.*, vol. 18, no. 3, pp. 667–670, 1979.

[19] G. Dharmadurai and N. S. S. Murthy, "A simplified expression for the minimum hotspot current in long, thin-film superconductors," *J. Low Temp. Phys.*, vol. 37, no. 3–4, pp. 269–276, Nov. 1979.

[20] G. Dharmadurai, "Self-heating-induced phenomena in long thin-film superconductors and their applications," *Phys. Status Solidi*, vol. 62, no. 1, pp. 11–33, 1980.

[21] A. G. Moshe, E. Farber, and G. Deutscher, "Optical conductivity of granular aluminum films near the Mott metal-to-insulator transition," *Phys. Rev. B*, vol. 99, no. 22, pp. 1–7, 2019.

[22] W. A. Little, "The Transport of Heat Between Dissimilar Solids at Low Temperatures," *Can. J. Phys.*, vol. 37, 1959.

[23] F. C. Wellstood, C. Urbina, and J. Clarke, "Hot-electron effects in metals," *Phys. Rev. B*, vol. 49, no. 9, pp. 5942–5955, 1994.

[24] J. K. Viljas and T. T. Heikkilä, "Electron-phonon heat transfer in monolayer and bilayer graphene," *Phys. Rev. B - Condens. Matter Mater. Phys.*, vol. 81, no. 245404, pp. 1–9, 2010.

[25] J. T. Karvonen, L. J. Taskinen, and I. J. Maasilta, "Observation of disorder-induced

weakening of electron-phonon interaction in thin noble-metal films," *Phys. Rev. B - Condens. Matter Mater. Phys.*, vol. 72, no. 1, pp. 1–4, 2005.

[26] G. Fagas *et al.*, "Lattice dynamics of a disordered solid-solid interface," *Phys. Rev. B*, vol. 60, no. 9, pp. 6459–6464, 1999.

[27] E. T. Swartz and R. O. Pohl, "Thermal boundary resistance," *Rev. Mod. Phys.*, vol. 61, no. 3, pp. 605–668, 1989.

[28] M. W. Brenner, D. Roy, N. Shah, and A. Bezryadin, "Dynamics of superconducting nanowires shunted with an external resistor," *Phys. Rev. B*, vol. 85, no. 22, p. 224507, Jun. 2012.

[29] D. Hazra, L. M. A. Pascal, H. Courtois, and A. K. Gupta, "Hysteresis in superconducting short weak links and μ-SQUIDs," *Phys. Rev. B - Condens. Matter Mater. Phys.*, vol. 82, no. 18, pp. 1–10, 2010.

[30] N. Kumar, T. Fournier, H. Courtois, C. B. Winkelmann, and A. K. Gupta, "Reversibility Of Superconducting Nb Weak Links Driven By The Proximity Effect In A Quantum Interference Device," *Phys. Rev. Lett.*, vol. 114, no. 15, pp. 1–5, 2015.

[31] F. W. J. Hekking, A. O. Niskanen, and J. P. Pekola, "Electron-phonon coupling and longitudinal mechanical-mode cooling in a metallic nanowire," *Phys. Rev. B - Condens. Matter Mater. Phys.*, vol. 77, no. 3, pp. 1–4, 2008.

[32] W. Frick, D. Waldmann, and W. Eisenmenger, "Phonon emission spectra of thin metallic films," *Appl. Phys.*, vol. 8, no. 2, pp. 163–171, 1975.

[33] J. Schleeh *et al.*, "Phonon black-body radiation limit for heat dissipation in electronics,"

*Nat. Mater.*, vol. 14, no. November 2014, pp. 187–192, 2015.

[34]  J. L. Rose, *Ultrasonic Guided Waves in Solid Media*, First. Cambridge University Press, 2014.

[35]  S. B. Kaplan, "Acoustic matching of superconducting films to substrates," *J. Low Temp. Phys.*, vol. 37, no. 3–4, pp. 343–365, Nov. 1979.

[36]  A. E. Dane, "Superconducting Photodetectors, Nanowires, and Resonators," Massachusetts Institute of Technology, 2019.

[37]  L. B. Wang, O.-P. Saira, D. S. Golubev, and J. P. Pekola, "Crossover between Electron-Phonon and Boundary-Resistance Limits to Thermal Relaxation in Copper Films," *Phys. Rev. Appl.*, vol. 12, no. 024051, p. 1, 2019.

[38]  A. J. Kerman, D. Rosenberg, R. J. Molnar, and E. a. Dauler, "Readout of superconducting nanowire single-photon detectors at high count rates," *J. Appl. Phys.*, vol. 113, no. 144511, 2013.

[39]  P. Li, P. M. Wu, Y. Bomze, I. V. Borzenets, G. Finkelstein, and A. M. Chang, "Retrapping current, self-heating, and hysteretic current-voltage characteristics in ultranarrow superconducting aluminum nanowires," *Phys. Rev. B*, vol. 84, no. 18, pp. 1–7, 2011.

[40]  S. B. Kaplan, C. C. Chi, D. N. Langenberg, J. J. Chang, S. Jafarey, and D. J. Scalapino, "Quasiparticle and phonon lifetimes in superconductors*," *Phys. Rev. B*, vol. 14, no. 11, pp. 4854–4873, 1976.

[41]  D. G. Cahill *et al.*, "Nanoscale thermal transport. II. 2003–2012," *Appl. Phys. Rev.*, vol. 1,

no. 1, p. 011305, Mar. 2014.

[42] M. Sidorova, "Timing Jitter and Electron-Phonon Interaction in Superconducting Nanowire Single-Photon Detectors ( SNSPDs )," 2020.

[43] S. Krause *et al.*, "Reduction of Phonon Escape Time for NbN Hot Electron Bolometers by Using GaN Buffer Layers," *IEEE Trans. Terahertz Sci. Technol.*, vol. 7, no. 1, pp. 1–7, 2017.

[44] P. J. de Visser, S. A. H. de Rooij, V. Murugesan, D. J. Thoen, and J. J. A. Baselmans, "Phonon-trapping enhanced energy resolution in superconducting single photon detectors," *arXiv:2103.06723v1*, pp. 1–14, 2021.

[45] T. A. Puurtinen, K. Rostem, P. J. de Visser, and I. J. Maasilta, "A Composite Phononic Crystal Design for Quasiparticle Lifetime Enhancement in Kinetic Inductance Detectors," *J. Low Temp. Phys.*, 2020.